\documentclass[sigconf]{acmart}
\AtBeginDocument{%
  }

\setcopyright{acmlicensed}
\copyrightyear{2025}
\acmYear{2025}
\acmDOI{XXXXXXX.XXXXXXX}
\acmConference[Conference acronym 'XX]{Make sure to enter the correct
  conference title from your rights confirmation email}{June 03--05,
  2018}{Woodstock, NY}
\acmISBN{978-1-4503-XXXX-X/2018/06}

\usepackage{multirow}

\begin{document}

\title{\textsc{CS-PaperSum}: A Large-Scale Dataset of AI-Generated Summaries for Scientific Papers}

\author{Javin Liu}
\email{javinliu@usc.edu}
\affiliation{%
  \institution{University of Southern California}
  \city{Los Angeles}
  \state{CA}
  \country{USA}
}

\author{Aryan Vats}
\email{aryanvat@usc.edu}
\affiliation{%
  \institution{University of Southern California}
  \city{Los Angeles}
  \state{CA}
  \country{USA}
}

\author{Zihao He}
\email{zihaoh@usc.edu}
\affiliation{%
  \institution{University of Southern California}
  \city{Los Angeles}
  \state{CA}
  \country{USA}
}

% \author{Dennis Perepech}
% \email{perepech@usc.edu}
% \affiliation{%
%   \institution{University of Southern California}
%   \city{Los Angeles}
%   \state{CA}

% \author{Aaron Ahmed}
% \email{arahmed@usc.edu}
% \affiliation{%
%   \institution{University of Southern California}
%   \city{Los Angeles}
%   \state{CA}

% \author{Yue Zhao}
% \email{yzhao010@usc.edu}
% \affiliation{%
%   \institution{University of Southern California}
%   \city{Los Angeles}
%   \state{CA}

\renewcommand{\shortauthors}{Trovato et al.}

\begin{abstract}
The rapid expansion of scientific literature in computer science presents challenges in tracking research trends and extracting key insights. Existing datasets provide metadata but lack structured summaries that capture core contributions and methodologies. We introduce \textsc{CS-PaperSum}, a large-scale dataset of 91,919 papers from 31 top-tier computer science conferences, enriched with AI-generated structured summaries using ChatGPT. To assess summary quality, we conduct embedding alignment analysis and keyword overlap analysis, demonstrating strong preservation of key concepts. We further present a case study on AI research trends, highlighting shifts in methodologies and interdisciplinary crossovers, including the rise of self-supervised learning, retrieval-augmented generation, and multimodal AI. Our dataset enables automated literature analysis, research trend forecasting, and AI-driven scientific discovery, providing a valuable resource for researchers, policymakers, and scientific information retrieval systems.\footnote{Our dataset is available at \url{https://github.com/zihaohe123/CS-PaperSum}.}
\end{abstract}



\keywords{Do, Not, Us, This, Code, Put, the, Correct, Terms, for,
  Your, Paper}

% \received{27 February 2025}

\maketitle

\section{Introduction}
The rapid expansion of scientific literature in computer science has created significant challenges for researchers attempting to stay informed about new developments, track emerging research trends, and synthesize insights from an overwhelming number of publications. Each year, thousands of papers are published in top-tier venues, spanning a wide range of subfields including artificial intelligence, natural language processing, computer vision, and cybersecurity. As the pace of research accelerates, the ability to efficiently extract key insights from the vast body of scientific literature has become increasingly important. Existing open-access repositories and metadata-rich datasets such as the Semantic Scholar Open Research Corpus (S2ORC) \cite{lo2020s2orc} and Microsoft Academic Graph (MAG) \cite{wang2020microsoft} provide bibliographic information, abstracts, and citation networks, but they lack structured summaries that capture the core contributions, methodologies, and future research directions of individual papers. This limitation makes large-scale automated analysis of research progress difficult, hindering the development of AI-driven tools for literature synthesis and trend detection.

To address these challenges, we introduce \textsc{CS-PaperSum}, a large-scale dataset of 91,919 computer science papers drawn from 31 major conferences. Unlike previous datasets that primarily provide raw text and metadata, \textsc{CS-PaperSum} incorporates structured AI-generated summaries produced using GPT-3.5 \cite{ouyang2022training}. Each summary distills the paper’s key contributions, novel methodologies, evaluation metrics, and future research directions, offering a concise yet comprehensive view of its content. By standardizing this information across a broad collection of papers, the dataset enables large-scale scientometric studies, automated literature reviews, and AI-assisted research synthesis.

The motivation for constructing this dataset arises from several critical challenges in scientific paper analysis. The sheer volume of research output makes it impractical for scholars to manually read and summarize papers at scale, even with the aid of search engines and citation networks \cite{fortunato2018science}. While platforms such as Google Scholar and Semantic Scholar index vast numbers of publications, they do not provide structured insights that help researchers quickly understand the significance of a paper beyond its title and abstract. Moreover, existing approaches to research trend analysis, such as topic modeling \cite{blei2003latent}, offer only high-level insights without capturing the fine-grained technical contributions of individual studies. Leveraging large language models to generate structured paper summaries, \textsc{CS-PaperSum} provides a scalable solution for extracting and organizing scientific knowledge.

To ensure the reliability of the AI-generated summaries, we conduct a thorough quality assessment, evaluating the semantic alignment between ChatGPT summaries and original papers. We employ SciBERT \cite{beltagy2019scibert}, a domain-specific transformer model trained on scientific text, to generate embeddings for both the original paper content and the AI-generated summaries. Visualization using t-SNE dimensionality reduction demonstrates that the embeddings of ChatGPT-generated summaries align well with those of the original papers while maintaining clear separation between different research domains. Additionally, we perform keyword overlap analysis using KeyBERT\footnote{https://maartengr.github.io/KeyBERT/} to compare the extracted keywords from the summaries and the original text. High overlap percentages indicate that the AI-generated summaries effectively preserve key concepts, making them a reliable tool for automated literature synthesis.

Beyond evaluating summary quality, we showcase a practical application of \textsc{CS-PaperSum} by analyzing research trends across major AI conferences. By extracting key topics from the structured summaries, we identify the evolution of research focus areas in NeurIPS, AAAI, EMNLP, and ICCV, revealing shifts in methodologies and emerging interdisciplinary intersections. For instance, we observe a transition in NeurIPS from traditional reinforcement learning methods \cite{mnih2015human} toward graph neural networks \cite{bronstein2021geometric} and large-scale self-supervised learning models \cite{liu2021self}. In NLP research, the dominance of statistical machine translation in earlier years \cite{koehn2003statistical} has been replaced by pretrained transformer-based architectures \cite{devlin2019bert}, retrieval-augmented generation \cite{lewis2020retrieval}, and efficient fine-tuning techniques \cite{houlsby2019parameter}. Computer vision research has similarly evolved, with ICCV showing increasing interest in vision transformers \cite{dosovitskiy2020image}, generative adversarial networks \cite{goodfellow2014generative}, and neural radiance fields for 3D reconstruction \cite{mildenhall2021nerf}. The dataset also allows for the detection of cross-disciplinary research trends, such as the adoption of self-supervised learning techniques originally developed for vision \cite{chen2020simple} into NLP applications \cite{gao2021simcse}.

% The broader implications of \textsc{CS-PaperSum} extend beyond research trend analysis. The dataset provides a valuable resource for training AI models in scientific document understanding, summarization, and automated review generation. It can support the development of AI-assisted literature recommendation systems, help funding agencies and policymakers track the evolution of scientific innovation, and enable more effective organization and retrieval of research knowledge. Furthermore, the dataset can serve as a benchmark for evaluating AI-generated scientific summaries, fostering advancements in fact-verifiable and citation-aware LLM-based summarization \cite{deyoung2021ms2}. Future work can explore improvements in AI-generated summaries by incorporating retrieval-augmented generation (RAG) \cite{lewis2020retrieval} and reinforcement learning with human feedback \cite{ouyang2022training} to enhance factual consistency and readability.

\textsc{CS-PaperSum} represents a significant step forward in the automated structuring and synthesis of scientific knowledge. By providing a large-scale dataset enriched with ChatGPT-generated structured summaries, we aim to bridge the gap between raw bibliographic data and actionable scientific insights. The dataset enables automated literature analysis, facilitates research trend detection, and provides a foundation for developing AI systems that assist in scientific discovery. As the landscape of AI and computer science continues to evolve, datasets like \textsc{CS-PaperSum} will play a crucial role in making scientific literature more accessible, navigable, and analyzable at scale.

\section{Related Work}
Our work builds upon prior efforts in scientific literature analysis, automatic summarization, and AI-driven research trend detection. 

Several large-scale datasets provide access to scientific papers and metadata, facilitating bibliometric studies and information retrieval. The Semantic Scholar Open Research Corpus (S2ORC) \cite{lo2020s2orc} contains full-text scientific papers and metadata, supporting NLP tasks such as citation analysis and topic modeling. Similarly, Microsoft Academic Graph (MAG) \cite{wang2020microsoft} provides a large knowledge graph of scholarly publications, including citation networks and author affiliations. Other datasets, such as CORE \cite{knoth2012core}, focus on aggregating open-access research papers. However, these resources primarily contain raw text, abstracts, and citation metadata, without structured summaries of research contributions, making large-scale literature synthesis difficult. \textsc{CS-PaperSum} addresses this gap by providing concise, AI-generated summaries that capture the main takeaways, methodologies, and future directions of each paper.

Automatic summarization of scientific papers has been widely studied in NLP, particularly in extractive and abstractive summarization. Early approaches used extractive methods, selecting key sentences based on salience and ranking \cite{louis2011automatic}. More recent advances leverage neural abstractive summarization, generating coherent summaries with transformers and pre-trained language models \cite{deyoung2021msˆ2}. SciBERT \cite{beltagy2019scibert} and SPECTER \cite{cohan2020specter} have been developed for scientific text processing, improving contextual representations of research papers. Our dataset extends this research by integrating ChatGPT-generated structured summaries, providing a new large-scale resource for evaluating LLM-based summarization in the scientific domain.

Recent studies have explored scientific trend analysis using citation networks and topic modeling. Chen et al. \citet{chen2017science} investigated the evolution of scientific fields using citation-based clustering, while Blei et al. \cite{blei2003latent} introduced Latent Dirichlet Allocation (LDA) for detecting emerging topics in research literature. Other works have examined research trends using co-authorship networks \cite{newman2001structure} and keyword co-occurrence analysis \cite{narong2023keyword}. However, these approaches rely primarily on citation graphs and word distributions rather than structured research summaries, limiting their ability to capture the nuances of evolving methodologies and experimental findings. \textsc{CS-PaperSum} enables fine-grained trend analysis by providing AI-generated summaries that explicitly highlight key contributions, experimental insights, and future research directions.

\section{Data}

\subsection{Dataset Collection and Content}
We collect open-access scientific papers using the Semantic Scholar API \cite{kinney2023semantic}, focusing on major computer science conferences spanning multiple subfields. After processing, our dataset comprises 91,919 papers from 31 conferences, covering diverse areas such as artificial intelligence, computer vision, natural language processing, computational theory, and computer architecture. A complete list of conferences\footnote{We include very few ACM conferences due to open access restrictions.} included in the dataset is provided in Table \ref{tab:confs}.

Our dataset spans from 2017 to 2024, offering a comprehensive longitudinal view of recent advancements in computer science research. The inclusion of papers from multiple years enables temporal trend analysis, making it possible to track the evolution of key topics, shifts in research focus, and the growth of different subfields.

For each paper, we extract and include the following metadata:  \textit{Title}, \textit{Authors}, \textit{Confrence}, \textit{year of publication}, \textit{abstract}, \textit{citation count}\footnote{Citation counts are dynamic and may change over time. Our dataset captures the values available at the time of collection.}.

The conclusion section extraction is performed using GROBID \cite{GROBID}, a machine learning-based tool for parsing scientific documents. We convert each paper’s PDF into structured XML format and identify the conclusion section by searching for the last enumerated section containing keywords such as “conclusion” or “discussion”. This approach ensures that we capture key takeaways from each paper, which are often useful for meta-analyses, automated summarization, and research trend forecasting.

% \usepackage{multirow}
\begin{table*}[ht]
\centering
\caption{A categorized list of the 31 conferences included in our dataset, organized by their respective research domains.}
% \addtolength{\tabcolsep}{-4.0pt}
\begin{tabular}{ll}
\hline
\multicolumn{1}{c}{\textbf{Research Subfield}}          & \multicolumn{1}{c}{\textbf{Conference}}                                       \\ \hline
\multirow{2}{*}{Computing in Biomedical Fields} &
  Medical Image Computing and Computer-Assisted Intervention (MICCAI) \\
                                                     & IEEE International Symposium on Biomedical Imaging (ISBI)                     \\ \hline
\multirow{2}{*}{Natural Language Processing}         & Empirical Methods in Natural Language Processing (EMNLP)                      \\
                                                     & Association for Computational Linguistics (ACL)                               \\ \hline
\multirow{2}{*}{Computer Vision}                     & IEEE Computer Vision and Pattern Recognition (CVPR)                           \\
                                                     & International Conference on Computer Vision (ICCV)                            \\ \hline
\multirow{2}{*}{Artificial Intelligence}             & AAAI Conference on Artificial Intelligence (AAAI)                             \\
                                                     & International Conference on Learning Representations (ICLR)                   \\ \hline
\multirow{3}{*}{Computational Theory}                & Conference and Workshop on Neural Information Processing Systems (NeurIPS)    \\
                                                     & International Conference on Artificial Intelligence and Statistics (AISTATS)  \\
                                                     & Annual Conference Computational Learning Theory (COLT)                        \\ \hline
\multirow{2}{*}{Computer Hardware \& Architecture}   & IEEE/ACM International Symposium on Microarchitecture (MICRO)                 \\
                                                     & IEEE International Symposium on High-Performance Computer Architecture (HCPA) \\ \hline
\multirow{3}{*}{Computer Networks \& Communications} & The Web Conference (WWW)                                                      \\
                                                     & Computer and Communications Security (CCS)                                    \\
                                                     & IEEE International Conference on Computer Communications (INFOCOM)            \\ \hline
\multirow{4}{*}{Computer Security \& Cryptography}   & Network and Distributed System Security Symposium (NDSS)                      \\
                                                     & USENIX Annual Technical Conference (USENIX)                                   \\
                                                     & Theory of Cryptography Conference (TCC)                                       \\
                                                     & Annual International Cryptology Conference (Crypto)                           \\ \hline
\multirow{2}{*}{Database \& Information Systems}     & Algorithms and Data Structures Symposium (WADS)                               \\
                                                     & Conference on Information and Knowledge Management (CIKM)                     \\ \hline
\multirow{2}{*}{Graphics and Computer-Aided Design} &
  IEEE International Conference on Automatic Face \& Gesture Recognition (FG) \\
                                                     & IEEE International Conference on Multimedia and Expo (ICME)                   \\ \hline
\multirow{2}{*}{Human Computer Interaction}          & Conference on Human Factors in Computing Systems (CHI)                        \\
                                                     & Human-Robot Interaction (HRI)                                                 \\ \hline
\multirow{2}{*}{Mixed and Augmented Reality}         & IEEE Virtual Reality Conference (VRIC)                                        \\
                                                     & International Symposium on Mixed and Augmented Reality (ISMAR)                \\ \hline
\multirow{3}{*}{Software Engineering}                & International Conference on Software Engineering (ICSE)                       \\
 &
  \begin{tabular}[c]{@{}l@{}}Architectural Support for Programming Languages \\ and Operating Systems Conference (ASPLOS)\end{tabular} \\
                                                     & Programming Language Design and Implementation (PLDI)                         \\ \hline
\end{tabular}
% \addtolength{\tabcolsep}{4.0pt}
\label{tab:confs}
\end{table*}

\subsection{Statistics}
\begin{figure}
    \centering
    \includegraphics[width=\linewidth]{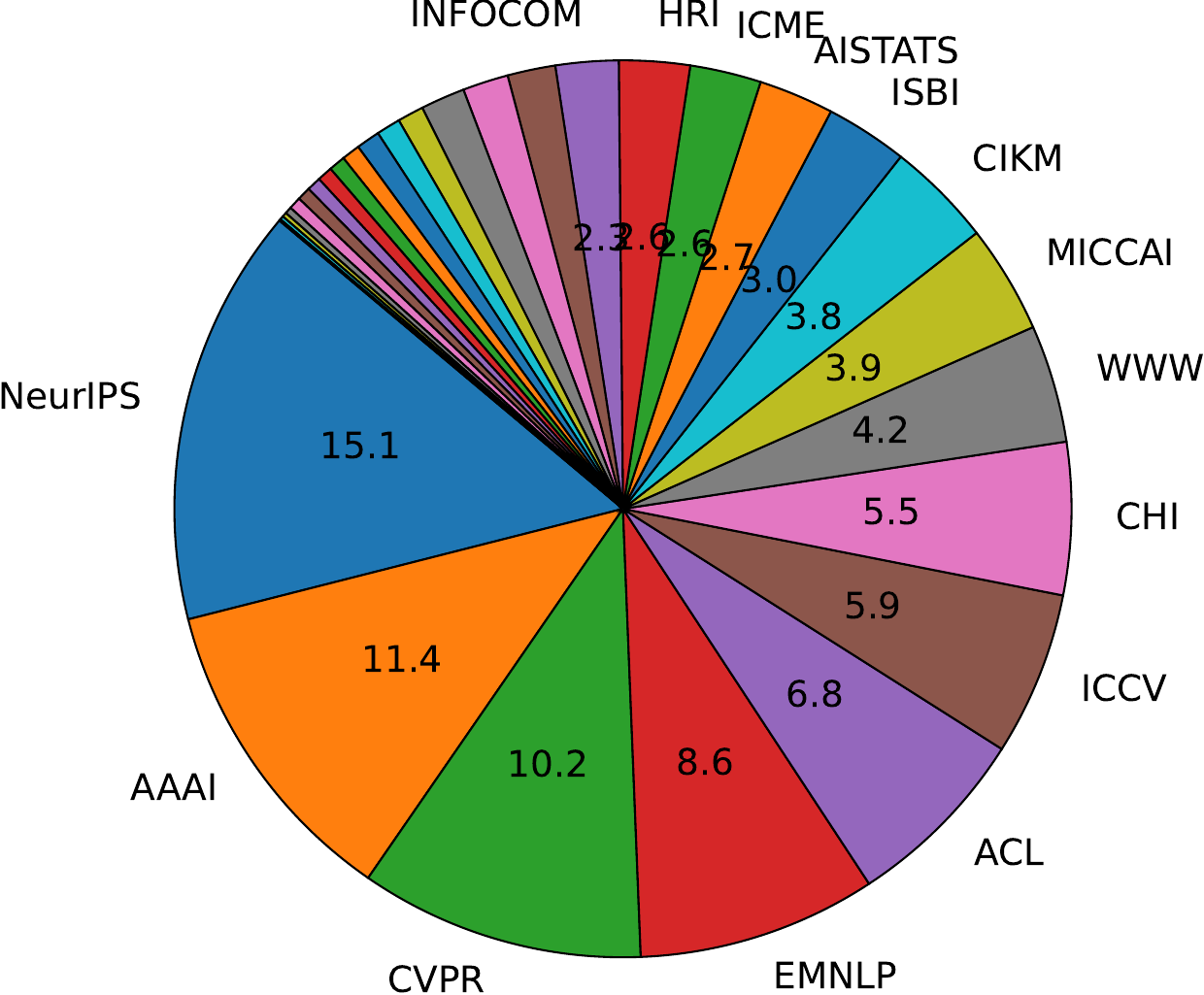}
    \caption{Fraction (\%) of papers published in each conference. Only conferences that contribute at least 2.3\% of the total publications are shown.
    }
    \label{fig:confs_pie}
\end{figure}

The distribution of papers per conference is visualized in Figure \ref{fig:confs_pie}, highlighting the relative contributions of different venues. The dataset includes papers from high-impact conferences such as NeurIPS, CVPR, ACL, EMNLP, ICCV, and AAAI, among others.

The yearly distribution of papers is detailed in Table \ref{tab:paper-count}, showing an overall increase in publications over time. This trend reflects the growing volume of research contributions in computer science, driven by advancements in AI, ML, and other subfields. Notably, from 2017 to 2023, the number of papers more than doubled, indicating an accelerating pace of research output.

\begin{table}[ht]
\centering
\addtolength{\tabcolsep}{-3.0pt}
\caption{Number of papers published per year from 2017 to 2024. The data illustrates a steady increase in research output over time.}
\begin{tabular}{ccccccccc}
\hline
\textbf{Year} &
  \textbf{2017} &
  \textbf{2018} &
  \textbf{2019} &
  \textbf{2020} &
  \textbf{2021} &
  \textbf{2022} &
  \textbf{2023} &
  \textbf{2024} \\ \hline
\textbf{Count} &
  8,120 &
  {8,704} &
  {12,642} &
  {13,130} &
  {14,851} &
  {15,491} &
  {18,461} &
  520 \\ \hline
\end{tabular}
\addtolength{\tabcolsep}{3.0pt}
\label{tab:paper-count}
\end{table}

\subsection{Productive Affiliations}
\begin{figure}[ht]
    \centering
    \includegraphics[width=\linewidth]{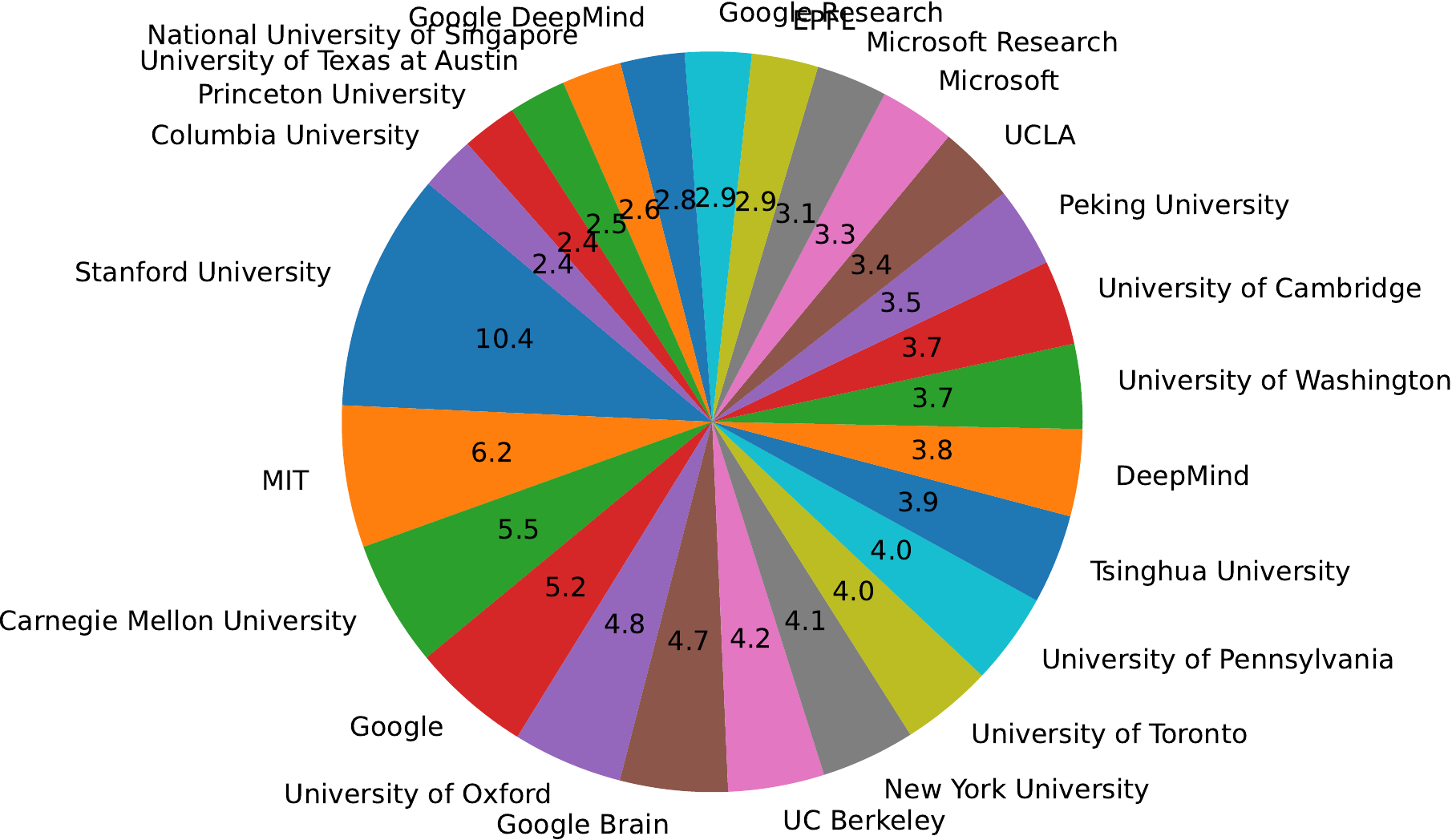}
    \caption{Fraction (\%) of papers published by the most productive affiliations. We focus on the leading universities, industry research labs, and institutions contributing to computer science research.}
    \label{fig:pie_chart_aff}
\end{figure}

Understanding the institutions that contribute the most to cutting-edge research is essential for assessing the global landscape of computer science research. In this study, we analyze the most productive affiliations by tracking the number of papers published by different universities, research labs, and industry organizations. Figure \ref{fig:pie_chart_aff} presents the top affiliations contributing to NeurIPS, one of the leading AI and machine learning conferences. The dataset also includes similar breakdowns for all 31 conferences, available in our GitHub repository.

From our analysis, we observe several key trends in the distribution of research productivity across institutions.

\paragraph{Dominance of elite research universities} Prestigious institutions such as MIT, Stanford, Carnegie Mellon University (CMU), UC Berkeley, University of Oxford, and the University of Toronto are consistently among the top contributors across multiple conferences. These universities have long-standing reputations for excellence in AI, computer vision, and natural language processing research.

\paragraph{Strong presence of industry labs} Industry research labs play a major role in shaping the field. Leading tech companies such as Google Research, DeepMind, Microsoft Research, Meta AI, and NVIDIA Research contribute significantly to top conferences. The presence of corporate research labs highlights the increasing commercialization of AI research and the close collaboration between academia and industry.

\paragraph{Regional distribution of top contributors} North America dominates, with U.S.-based institutions contributing the largest share of publications across conferences. The United States remains the hub for AI research, particularly in fields like deep learning and computer vision. Europe maintains a strong research presence, led by institutions such as Oxford, Cambridge, ETH Zurich, and INRIA. Asia has seen substantial growth in research output, particularly from Tsinghua University, Peking University, National University of Singapore, and the Chinese Academy of Sciences. This reflects China and Singapore’s increasing investment in AI research.

\subsection{Influential Conferences}
While publication volume provides insights into research productivity, the impact of research is best measured by citation influence. To assess conference influence, we use the median citation count of papers published at each venue. Table \ref{tab:influential_confs} presents the top 10 most impactful conferences based on this metric.

Our analysis reveals that conferences in artificial intelligence and computer vision, such as CVPR, ICLR, and ICCV, consistently rank among the most influential. CVPR, in particular, leads in citation impact, reflecting the rapid advancements and widespread applications of computer vision research in areas such as autonomous driving, medical imaging, and facial recognition. Similarly, ICLR has emerged as a top-tier venue for deep learning and representation learning, fueled by the growing importance of neural architectures and optimization techniques. The sustained impact of ICCV further highlights the strong research interest in visual understanding and scene recognition, which continues to push the boundaries of AI capabilities.

Beyond AI, our results indicate that conferences in security, programming languages, and system architecture also demonstrate substantial influence. NDSS and USENIX Security are among the highest-ranked venues, underscoring the increasing importance of cybersecurity, network protection, and adversarial AI defenses. Research presented at these conferences often leads to significant advancements in data privacy, secure computation, and cryptographic protocols, making them critical for both academia and industry. Similarly, PLDI and ASPLOS, which focus on programming languages and system design, maintain strong citation impact. The sustained influence of these conferences highlights the foundational role of compilers, parallel computing, and energy-efficient hardware architectures in modern computing.

One particularly interesting observation is that citation impact does not always correlate directly with conference volume. While high-volume AI conferences like NeurIPS and AAAI are widely regarded as prestigious, they do not always rank at the top in terms of median citation count. This suggests that while these venues facilitate a large number of research contributions, the influence of individual papers may be more variable compared to smaller, more selective conferences. In contrast, theoretical AI conferences, such as COLT, tend to have fewer but highly influential papers, reflecting the specialized and long-term impact of foundational theoretical work.

The factors contributing to a conference’s influence extend beyond the number of citations. Conferences covering highly applied and societally relevant topics, such as AI, cybersecurity, and large-scale computing systems, tend to attract broader readership and citation counts. Moreover, interdisciplinary conferences that bridge multiple fields, such as CHI and CVPR, often achieve high impact due to their relevance across industry, academia, and policymaking. The degree of industry participation also plays a significant role in conference influence. Venues such as NeurIPS, CVPR, and ICLR benefit from strong engagement from leading AI companies, startups, and technology firms, ensuring that the research presented is not only theoretically rigorous but also highly applicable to real-world scenarios.

Selectivity and acceptance rates further shape a conference’s influence. More competitive venues with rigorous peer review processes, such as ICLR and CVPR, tend to yield higher-impact papers, as only the most novel and technically sound research is accepted. Additionally, high-profile keynote speakers, invited talks, and industry sponsorships contribute to the visibility and influence of a conference. Prominent researchers presenting groundbreaking work can set new research agendas, leading to long-term academic and industrial adoption of novel ideas.

\begin{table}[ht]
\centering
\caption{Top 10 conferences ranked by their median citation count, reflecting their relative impact on the research community.}
\label{tab:influential_confs}
\begin{tabular}{cccccc}
\hline
\textbf{Conference} & \textbf{CVPR} & \textbf{ICLR}   & \textbf{ICCV}   & \textbf{NDSS} & \textbf{PLDI} \\
\textbf{Impact} & 34 & 25 & 20 & 19 & 16 \\ \hline
\textbf{Conference} & \textbf{CHI}  & \textbf{USENIX} & \textbf{ASPLOS} & \textbf{COLT} & \textbf{HPCA} \\
\textbf{Impact} & 16 & 15 & 14 & 14 & 13 \\ \hline
\end{tabular}
\end{table}

\section{Summarizing the Papers}

To enhance the usability and accessibility of our dataset, we generate structured summaries of each paper using ChatGPT (GPT-3.5) \cite{ouyang2022training}. While traditional datasets primarily provide raw text and metadata \cite{lo2020s2orc, wang2020microsoft, knoth2012core}, our dataset, \textsc{CS-PaperSum}, incorporates automated summaries that distill key insights from each paper, including its main contributions, methodological innovations, and future directions. These structured summaries serve as concise yet informative representations of scientific papers, aiding researchers in quickly grasping the key findings of a paper without having to read the full text.

The motivation for incorporating ChatGPT-generated summaries is twofold. First, the rapid growth in computer science research output makes it increasingly challenging for researchers to stay updated with recent advancements. By providing structured summaries, we facilitate efficient information retrieval and enable large-scale trend analysis. Second, while existing datasets provide metadata such as titles, abstracts, and citation counts, they often lack standardized key takeaways that could be leveraged for downstream tasks such as research trend forecasting, automated literature reviews, and AI-assisted scientific discovery.

To generate the summaries, we prompt ChatGPT with the title, abstract, and conclusion of each paper and instruct it to extract key insights in a structured list format. The prompt is designed to ensure that the model captures not only the main contributions of the paper but also its broader impact and future research directions. Below is the structured output format used for the summaries. \\

\noindent \textit{You will help summarize research. Given an paper title and abstract summarize the key takeaways, importance of the paper, what does the paper bring to the table that was unique from previous works, identify and list the key topics, models, and techniques discussed. Focus on condensing these elements into a concise list format. Each list item should represent a distinct topic or technique mentioned in the paper which is one to two words, any future works, performance measures (f1 score, accuracy, etc of the work /other works mentioned), performance effectiveness with 3 possible values highly effective not effective or moderately effective, future works, sentiments for the models/techniques, score for paper influence between 0 to 10. Return in the following format.}\\

\noindent \textit{Key Takeaways: xxx}\\
\noindent \textit{Importance: xxx}\\
\noindent \textit{Model/Method Proposed: xxx}\\
\noindent \textit{Performance: xxx}\\
\noindent \textit{Effectiveness: xxx}\\
\noindent \textit{Future Work: xxx}\\

\noindent \textit{For Key Takeaways, Importance, and Model/Method Proposed, please have at least 5 points in bullet point format. }\\

\noindent \textit{Title: [title]}\\
\noindent \textit{Abstract: [abstract]}\\
\noindent \textit{Conclusion: [conclusion]}\\

By structuring the summaries in this format, we provide a standardized representation of each paper, making it easier to compare research across different conferences and subfields. Given the importance of this additional information, we name our dataset \textsc{CS-PaperSum}, reflecting its core goal of summarizing computer science research papers using ChatGPT-generated key insights.

\subsection{Quality Assessment of Summaries}
To ensure the reliability and fidelity of the ChatGPT-generated summaries, we conduct both embedding alignment analysis and keyword overlap analysis. These evaluations assess whether the summaries accurately preserve the semantic and topical structure of the original papers.

\paragraph{Embedding Alignment Analysis}

\begin{figure*}[ht]
    \centering
    \includegraphics[width=0.8\linewidth]{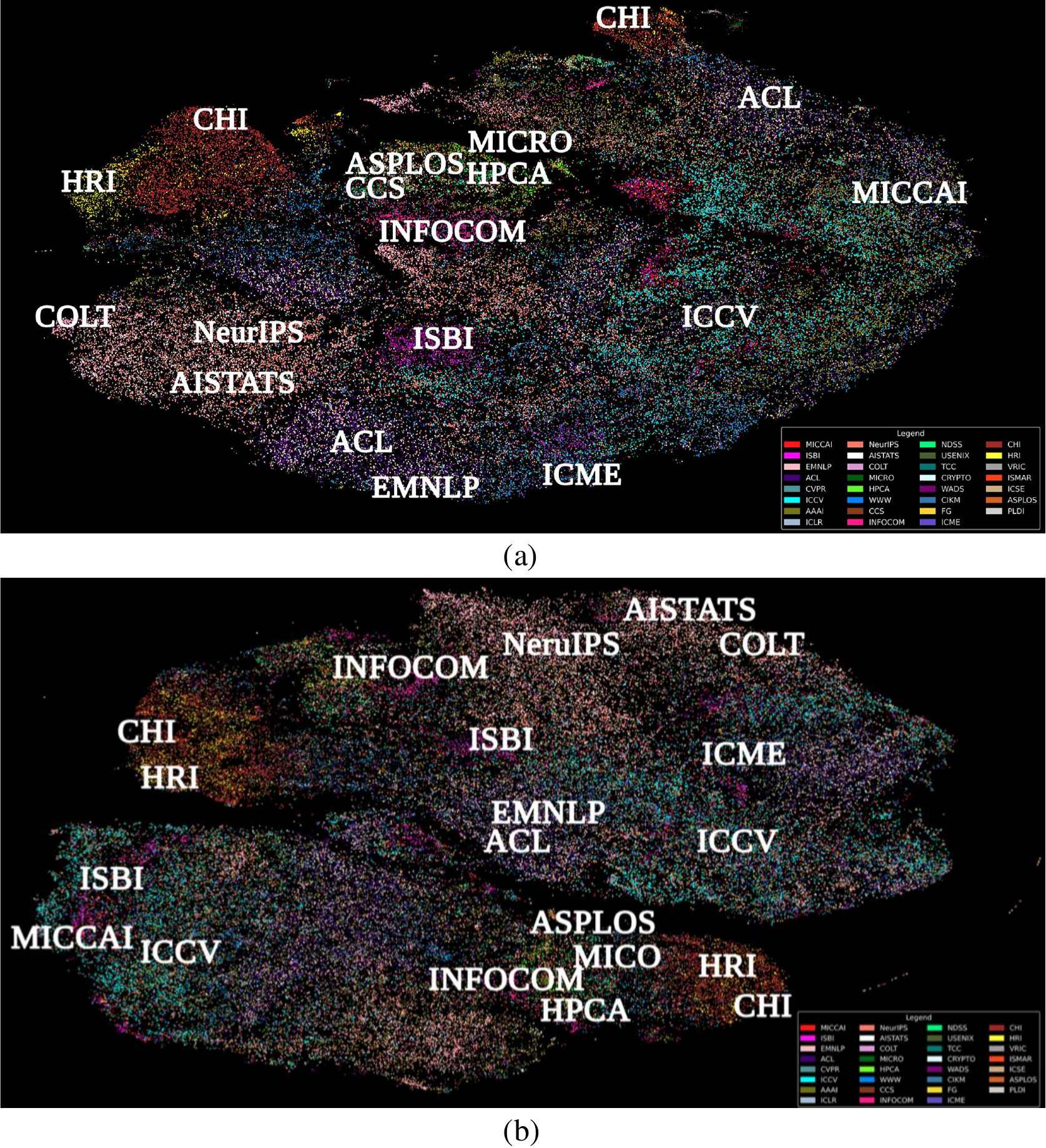}
    \caption{t-SNE visualization of paper embeddings based on (a) the original paper content, including the title, abstract, and conclusion, and (b) the ChatGPT-generated summaries (``Key Takeaways''). Each point represents a paper, and different colors indicate different conferences. The spatial clustering patterns suggest that the AI-generated summaries effectively preserve the semantic structure of the original papers while maintaining distinctions between research domains.}
    \label{fig:tsne}
\end{figure*}

We analyze how well the ChatGPT summaries retain the semantic characteristics of the original papers using embedding-based visualization. Specifically, we use SciBERT \cite{beltagy2019scibert}, a domain-specific BERT model pretrained on scientific literature, to obtain vector representations of both the original papers and their corresponding summaries. We then apply t-SNE dimensionality reduction to project these high-dimensional embeddings into a 2D space, as illustrated in Figure \ref{fig:tsne}.

Our visualization demonstrates that the embeddings of summaries are closely aligned with those of the original papers, while still maintaining clear conference-level clustering. This suggests that ChatGPT effectively captures the distinctive characteristics of different research domains, preserving key topic structures across conferences such as NeurIPS (AI/ML), CVPR (Computer Vision), and ACL (NLP).

\paragraph{Keyword Overlap Analysis}
To further evaluate the summaries, we measure keyword overlap between the original papers and their ChatGPT-generated summaries. We use KeyBERT\footnote{https://maartengr.github.io/KeyBERT/}, a BERT-based keyword extraction tool, to identify the top 10 most relevant keywords for each paper. These keywords are extracted based on their semantic similarity to the full text, ensuring that they represent the most salient concepts discussed in the paper.

\begin{figure}[ht]
    \centering
    \includegraphics[width=0.9\linewidth]{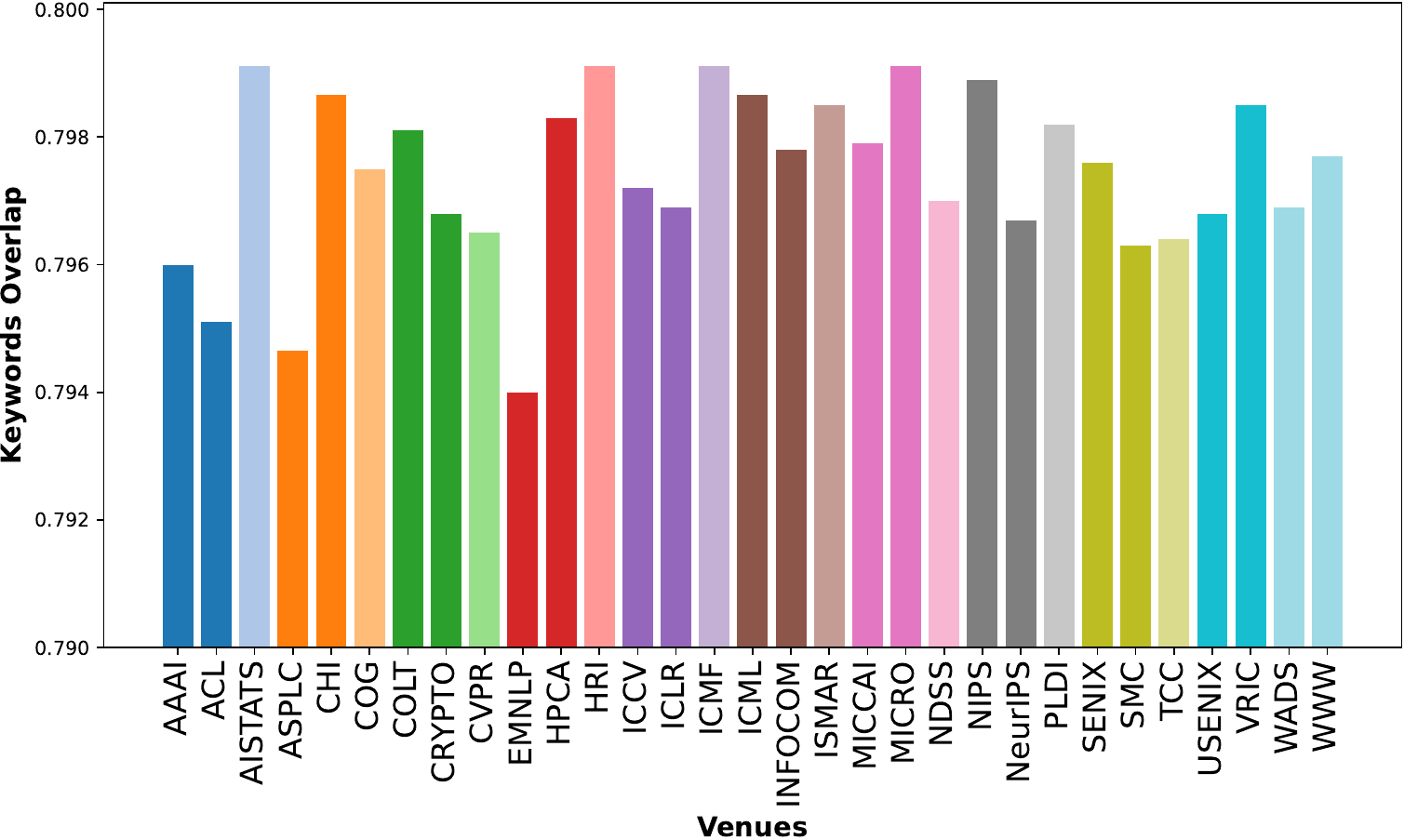}
    \caption{Keyword overlap between original papers and their ChatGPT-generated summaries (“Key Takeaways”) across different conferences. The overlap is measured using KeyBERT-based keyword extraction, where higher values indicate greater retention of key concepts in the summaries. The results demonstrate that the AI-generated summaries effectively capture the main topics of the original papers while varying slightly across different research domains.}
    \label{fig:keywords_overlap}
\end{figure}

For each paper, we compute the intersection of keywords between the original document and its corresponding summary, quantifying how well the generated summary preserves key topics. The results, presented in Figure \ref{fig:keywords_overlap}, show a high degree of overlap, particularly in AI, machine learning, and natural language processing conferences. This indicates that ChatGPT summaries effectively retain core concepts and terminology, making them a reliable tool for summarizing scientific literature.

\subsection{Case Study: Identifying Research Trends}

One of the most compelling applications of \textsc{CS-PaperSum} is its ability to uncover evolving research trends across different subfields of computer science. By analyzing the ChatGPT-generated summaries, we can systematically identify emerging topics, shifting methodological paradigms, and evolving focus areas within major conferences. In this section, we demonstrate how our dataset facilitates trend analysis by examining four major AI-related conferences: NeurIPS, AAAI, EMNLP, and ICCV, which represent machine learning, artificial intelligence, natural language processing, and computer vision, respectively.

\paragraph{Temporal Evolution of Research Topics}

Having structured summaries enables longitudinal analysis of research priorities. By extracting the most frequently occurring keywords from the “Key Takeaways” section of each paper, we track the prominence of different topics over time. Table \ref{tab:keywords} presents the dominant research themes in NeurIPS, AAAI, EMNLP, and ICCV.

At NeurIPS, research initially focused on reinforcement learning (RL) and optimization methods, but in recent years, the field has shifted towards graph neural networks (GNNs) \cite{bronstein2021geometric}, self-supervised learning \cite{liu2021self}, and large-scale generative models \cite{brown2020language}. Additionally, concerns about AI robustness have driven extensive research into adversarial learning \cite{goodfellow2014generative} and uncertainty quantification.

In AAAI, research has traditionally covered a broad range of AI subfields, including multi-agent systems, evolutionary computation, and automated planning. However, a significant shift towards trustworthy AI is now evident, with a growing number of papers on human-AI collaboration \cite{shneiderman2020human}, and fairness in AI systems \cite{mehrabi2021survey}. The increasing focus on AI governance and ethical considerations suggests that the field is adapting to the regulatory landscape surrounding AI deployment \cite{jobin2019global}.

EMNLP has undergone one of the most profound transformations due to the rapid advancements in deep learning-based NLP models. Earlier years featured substantial work on statistical machine translation (SMT) \cite{koehn2003statistical} and linguistic feature engineering \cite{manning1999foundations}. However, the rise of large-scale pretrained transformers \cite{devlin2019bert} has led to a paradigm shift toward self-supervised learning, retrieval-augmented generation (RAG) \cite{lewis2020retrieval}, and efficient fine-tuning methods \cite{houlsby2019parameter}. The emergence of prompt engineering \cite{liu2023visual} and instruction tuning \cite{wei2022finetuned} highlights the increasing interest in making language models more controllable and adaptable.

At ICCV, research has evolved from classic object detection \cite{ren2015faster} and scene segmentation \cite{long2015fully} to incorporating self-supervised learning techniques \cite{caron2021emerging} and transformers for vision tasks \cite{dosovitskiy2020image}. Recent breakthroughs in Neural Radiance Fields (NeRFs) \cite{mildenhall2021nerf} and diffusion-based generative models \cite{ho2020denoising} indicate a shift towards high-fidelity image and video synthesis, with applications in virtual reality, computational photography, and 3D reconstruction.

\paragraph{Cross-Conference Intersections and Emerging Interdisciplinary Areas}

Beyond tracking individual conference trends, \textsc{CS-PaperSum} allows us to examine cross-disciplinary influences within AI research. A significant finding from our dataset is the increasing overlap between research fields, where ideas originally developed in one domain are adopted by another.

One example is the widespread adoption of self-supervised learning (SSL), initially popularized in computer vision (ICCV, CVPR) \cite{grill2020bootstrap}, which has now become a dominant paradigm in natural language processing (EMNLP, ACL) \cite{liu2021self}. SSL has enabled the development of models such as SimCLR \cite{chen2020simple} and BYOL \cite{grill2020bootstrap}, which have inspired new approaches for contrastive pretraining in NLP \cite{gao2021simcse}.
Similarly, reinforcement learning (RL), once primarily associated with robotics and game-playing AI \cite{silver2017mastering}, has increasingly been applied to NLP and structured generation tasks. 

Another notable intersection is between symbolic reasoning and deep learning, a hybrid approach known as Neuro-Symbolic AI \cite{garcez2019neural}. This paradigm is gaining traction in AAAI and NeurIPS, as it aims to bridge the gap between deep learning’s pattern recognition capabilities and the logical reasoning strengths of traditional AI \cite{manhaeve2021neural}.

\paragraph{Implications for the Future of AI Research}

Our dataset also reveals potential future directions in AI research based on emerging and declining topics across conferences. The growing focus on scalability and efficiency suggests that research will continue to explore parameter-efficient architectures \cite{dettmers2023qlora}, energy-efficient AI \cite{schwartz2020green}, and federated learning \cite{kairouz2021advances} to address the increasing computational demands of large-scale models.

Another critical research direction is AI alignment and safety, as evidenced by the rise of work on bias mitigation, robustness, and explainability \cite{weidinger2022taxonomy, chen2024susceptible, he2024community, he2024whose, nguyen2025smoothing, he2024reading}. With the deployment of AI in high-stakes applications (e.g., healthcare, finance, and law) \cite{chu2024improving}, the need for transparent, fair, and accountable AI systems will likely become a central theme in future research \cite{amodei2016concrete}.

The expansion of multimodal learning, where models process multiple types of inputs such as text, images, and audio, is another promising area \cite{baltruvsaitis2018multimodal}. The development of vision-language models \cite{radford2021learning} and cross-modal retrieval techniques \cite{alayrac2022flamingo, he2023alcap} suggests a growing trend towards AI systems that can integrate information from multiple modalities to perform more human-like reasoning.

\begin{table}[ht]
\centering
\caption{Top keywords extracted from major AI conferences, highlighting the dominant research topics in NeurIPS, AAAI, EMNLP, and ICCV. The keywords reflect recurring themes in each conference, such as reinforcement learning and adversarial training in NeurIPS, multi-agent systems and evolutionary learning in AAAI, language modeling and text generation in EMNLP, and computer vision tasks like inpainting and style transfer in ICCV. These keywords provide insights into the evolving focus areas of each research community.}
\label{tab:keywords}
\begin{tabular}{cl}
\hline
\textbf{Venue} &
  \multicolumn{1}{c}{\textbf{Top Keywords}} \\ \hline
NeurIPS &
  \begin{tabular}[c]{@{}l@{}}adversarial, adversarial training\\ neural networks, graph neural networks\\ obejct detection, translation\\ reinforce learning\\ gradient descent, stochastic\\ bandit, contextual bandits, sampling\end{tabular} \\ \hline
AAAI &
  \begin{tabular}[c]{@{}l@{}}agent, multi agent, multi objective\\ multitasking, evolutionary multitasking\\ reinforce learning, domain adaptation\\ adversarial training, attacks\\ graph neural, knowledge\end{tabular} \\ \hline
EMNLP &
  \begin{tabular}[c]{@{}l@{}}machine translation, neural machine\\ text generation, generation\\ language models, large language models\\ word embeddings, word representations\\ relation extraction, event extraction\\ knowledge graph, commonsense\\ dialogue, summarization, abstractive dialogue\\ semantic analysis, multimodal sentiment\\ question answering, qa\end{tabular} \\ \hline
ICCV &
  \begin{tabular}[c]{@{}l@{}}inpainting, video inpainting\\ pose, point, domain\\ style transfer, style, attention, person identification\\ text recognition, scene text, text image\\ adversarial, generative adversarial, metric learning\end{tabular} \\ \hline
\end{tabular}
\end{table}

% \section{Potential Applications and Future Directions}

% \textsc{CS-PaperSum} opens up numerous possibilities for research and practical applications, such as

% \begin{itemize}
%     \item Automated Literature Review Assistant: Develop tools to quickly synthesize research on specific topics, aiding researchers in literature reviews.
%     \item Cross-Domain Innovation Catalyst: Identify potential applications of techniques from one domain to another, fostering interdisciplinary research.
%     \item Trend Prediction and Research Gap Identification: Analyze evolving topics and methodologies to forecast future research trends and identify underexplored areas.
%     \item Personalized Research Recommendation System: Create systems that suggest relevant papers and potential collaborators based on a researcher's interests and past work.
%     \item Enhanced Scientific Search Engine: Develop more context-aware search tools that understand the key contributions of papers beyond simple keyword matching.
%     \item Curriculum Development and Educational Tools: Create up-to-date course materials and interactive learning modules reflecting current research trends.
%     \item Research Impact Analysis: Develop new metrics for assessing research impact by tracing the spread of concepts across papers and conferences over time.
% \end{itemize}

% Future work could focus on extending the dataset to include more conferences and longer time spans, improving the summarization process, and developing benchmarks for various tasks using this dataset.

\section{Conclusion}
In this work, we introduced \textsc{CS-PaperSum}, a large-scale dataset of 91,919 computer science papers from 31 top-tier conferences, enriched with ChatGPT-generated structured summaries. Unlike existing datasets that provide only metadata and raw text, \textsc{CS-PaperSum} offers standardized summaries capturing key takeaways, methodologies, evaluation metrics, and future research directions. Through embedding alignment analysis and keyword overlap evaluation, we demonstrated that the AI-generated summaries effectively preserve the semantic and topical characteristics of the original papers. Furthermore, our case study on AI research trends revealed evolving methodological paradigms and interdisciplinary connections, highlighting the dataset’s potential for large-scale scientometric studies.

Beyond research trend analysis, \textsc{CS-PaperSum} opens several promising avenues for future work. First, it can serve as a benchmark for AI-driven scientific summarization, allowing comparisons between LLM-generated summaries and human-written abstracts to evaluate factual accuracy and coherence. Second, the dataset can facilitate the development of automated literature review tools, enabling researchers to quickly synthesize insights across large collections of papers. Third, integrating retrieval-augmented generation (RAG) could improve AI-assisted scientific question-answering and information retrieval systems. Additionally, multi-document summarization models could leverage this dataset to generate comparative analyses of research across different papers or conferences.

\begin{acks}
We thank Dennis Perepech and Aryan Vats for their help.
\end{acks}

\bibliographystyle{ACM-Reference-Format}
\bibliography{sample-base,custom}

\appendix

\end{document}